\documentclass{aa}  
\bibpunct{(}{)}{;}{a}{}{,} 
\usepackage{natbib}
\usepackage{bigstrut}
\usepackage{graphicx}
\usepackage{txfonts}
\def\o17{$^{17}$O}                %
\def\orat{$^{16}$O/$^{17}$O}
\def\rate{$^{17}$O$(p,\alpha)^{14}$N}

\begin{document}

   \title{The impact of the revised $^{17}$O$(p,\alpha)^{14}$N reaction rate on  $^{17}$O stellar abundances and yields} 
   \titlerunning{The impact of the revised $^{17}$O$(p,\alpha)^{14}$N} 

   \author{O. Straniero\inst{1,2}
           \and
           C.G.Bruno\inst{5}
           \and           
           M. Aliotta\inst{5}
           \and
           A. Best\inst{6}
           \and
           A. Boeltzig\inst{3} 
           \and 
           D. Bemmerer\inst{4}           
           \and
           C. Broggini \inst{7}
           \and
             A. Caciolli\inst{7,8}
           \and
             F. Cavanna\inst{9}
           \and 
             G.F. Ciani\inst{3}             
           \and
             P. Corvisiero\inst{9}
           \and
             S. Cristallo\inst{1,16}
           \and 
             T. Davinson\inst{5}   
           \and
             R. Depalo\inst{7,8}
           \and  
             A. Di Leva\inst{6}
           \and
             Z. Elekes\inst{10}
           \and
             F. Ferraro\inst{9}
           \and  
             A. Formicola\inst{2}
           \and
             Zs. F\"{u}l\"{o}p\inst{10}
           \and
              G. Gervino\inst{11}
           \and  
              A. Guglielmetti\inst{12} 
           \and
              C. Gustavino\inst{13}
           \and
              G. Gy\"{u}rky\inst{10}
           \and   
              G. Imbriani\inst{6}
           \and    
              M. Junker\inst{2}
           \and
              R. Menegazzo\inst{7}
           \and   
              V. Mossa\inst{14}
           \and
              F.R. Pantaleo\inst{14}
           \and
              D. Piatti\inst{7,8} 
           \and
              L. Piersanti\inst{1,16}        
           \and
              P. Prati\inst{9}
           \and
              E. Samorjai\inst{10}
           \and
              F. Strieder\inst{15}
           \and   
             T. Sz"{u}cs\inst{4} 
           \and
              M.P. Tak\'{a}cs\inst{4}   
           \and
             D. Trezzi\inst{11}
}        

   \institute{
           INAF, Osservatorio Astronomico di Teramo, 64100 Teramo, Italy\footnote{\email{straniero@oa-teramo.inaf.it}}
        \and
           INFN, Laboratori Nazionali del Gran Sasso (LNGS), 67100 Assergi, Italy
        \and
           Gran Sasso Science Institute, INFN, Viale F. Crispi 7, 67100 L'Aquila, Italy   
        \and
           Helmholtz-Zentrum Dresden-Rossendorf, Bautzner Landstr 400, 01328 Dresden, Germany
        \and
           SUPA, School of Physics and Astronomy, University of Edinburgh, EH9 3FD Edinburgh, United Kingdom      
        \and
           Universit\`a di Napoli "Federico II" and INFN, Sezione di Napoli, 80126 Napoli, Italy
        \and
           INFN, Sezione di Padova, Via Marzolo 8, 35131 Padova, Italy
        \and
            Department of Physics and Astronomy, University of Padova, Via Marzolo 8, 35131 Padova, Italy  
        \and
             Universit\`a degli Studi di Genova and INFN, Sezione di Genova, Via Dodecaneso 33, 16146 Genova, Italy
        \and
            Institute for Nuclear Research (MTA ATOMKI), PO Box 51, HU-4001 Debrecen, Hungary
        \and
           Universit\`a degli Studi di Torino and INFN, Sezione di Torino, Via P. Giuria 1, 10125 Torino, Italy
        \and 
           Universit\`a degli Studi di Milano and INFN, Sezione di Milano, Via G. Celoria 16, 20133 Milano, Italy
        \and
           INFN, Sezione di Roma La Sapienza, Piazzale A. Moro 2, 00185 Roma, Italy
        \and
           Universit\`a degli Studi di Bari and INFN, Sezione di Bari, 70125 Bari, Italy
        \and
           South Dakota School of Mines, 501 E. Saint Joseph St., SD 57701 USA 
        \and
           INFN, Sezione di Perugia, 06123 Perugia, Italy  
}

   \date{Received ; accepted }

 
  \abstract
   {Material processed by the CNO cycle in stellar interiors is enriched in $^{17}$O. 
   When mixing processes from the stellar surface reach these layers, as occurs when 
   stars become
   red giants and undergo the first dredge up, the abundance of $^{17}$O increases. 
   Such an occurrence explains the drop of the \orat~ observed in  RGB stars
   with mass larger than $\sim1.5$ M$_\odot$.
    As a consequence, the interstellar medium is 
   continuously polluted by the wind of evolved stars enriched in \o17~.}
   {Recently, the  Laboratory for Underground Nuclear Astrophysics (LUNA) collaboration 
    released an improved rate of the \rate~ reaction. 
   In this paper we discuss the impact that the revised rate has on the $^{16}$O/$^{17}$O ratio 
   at the stellar surface and on  \o17~ stellar yields.}
   {We computed stellar models of initial mass between 1 and 20 M$_\odot$ and compared the results 
   obtained by adopting the revised rate of the \rate~ 
   to those obtained using previous rates. }
   {The post-first dredge up $^{16}$O/$^{17}$O ratios are about 20\% larger than previously obtained.
   Negligible variations are found in the case of the second and the third dredge up. In spite of the 
   larger \rate~ rate, we confirm previous claims that an extra-mixing process on the red giant branch, 
   commonly invoked to explain the low carbon isotopic ratio observed in bright 
   low-mass giant stars, marginally affects the \orat~ ratio. Possible 
   effects on AGB extra-mixing episodes are also discussed. 
   As a whole, a substantial reduction of  \o17~ stellar yields is found. 
   In particular, the net yield of stars with mass ranging  between 2 and 20 M$_\odot$ 
   is 15 to 40\% smaller than previously estimated.}
   {The revision of the \rate~ rate has a major impact on the interpretation of the \orat~ observed in
   evolved giants, in stardust grains and on the \o17~ stellar yields.}


   \maketitle
%

\section{Introduction}
The two lighter oxygen isotopes participate in the NO section of the CNO cycle.
When CNO reaches an equilibrium, as occurs in the 
H-burning shell of a giant star or in the central region of main-sequence stars,
the $^{16}$O/$^{17}$O ratio is simply given by the inverse ratio of the respective proton-capture 
reaction rates,
\begin{eqnarray}
 &\displaystyle\frac{^{16}\rm{O}}{^{17}\rm{O}}=\frac{(<\sigma v>_{^{17}\rm{O}(p,\gamma)^{18}\rm{F}}+
 <\sigma v>_{^{17}\rm{O}(p,\alpha)^{14}\rm{N}})}{<\sigma v>_{^{16}\rm{O}(p,\gamma)^{17}\rm{F}}}&
.\end{eqnarray}
In practice, since the three reaction rates only depend on temperature, 
this isotopic ratio traces the temperature profile within the H-burning zone. 
Variations of the $^{16}$O/$^{17}$O
isotopic ratio appear at
the surface when the ashes of the H burning are mixed with the stellar surface as a result of, for
instance, convective instabilities affecting the stellar envelope in red
giant (RGB) and asymptotic giant (AGB) stars. This isotopic ratio can be measured from the 
IR spectra of giant stars. In early studies, the CO vibration-rotation, 4.6 $\mu m$ fundamental or 2.3 $\mu m$
first overtone bands were commonly used \citep{harris1988, smith1990}. 
With the advent of IR spectrographs characterized by 
high dispersion and high signal-to-noise ratios, more lines in the near-IR spectrum can be exploited to derive the CNO 
isotopic composition of giant stars  \citep{garcia2010,abia2012,lebzelter2015,hinkle2016}. 
These measurements allow us
to probe the depth attained by the mixing episodes that are responsible for the
chemical variations observed at the stellar surface. We recall that in addition
to convection, other processes may induce  or contribute to deep mixing
episodes, such as rotation, magnetic buoyancy, gravity waves, and thermohaline circulation \citep{dearborn1992, boothroyd1994,
eleid1994, charbonnel1995, nollet2003, denissenkov2003, busso2007, eggleton2008, palmerini2011}. 
In all cases, the reliability of the deep mixing probe depends on our knowledge of the
rates of the nuclear reactions determining the  $^{16}$O$/^{17}$O ratio in the H-burning
regions, i.e, $^{16}$O$(p,\gamma)^{17}$F, $^{17}$O$(p,\gamma)^{18}$F and  
$^{17}$O$(p,\alpha)^{14}$N. The
last two reactions have recently been studied by the LUNA collaboration
\citep{scott2012,dileva2014,bruno2016} to obtain precise underground measurements of their cross sections at
relatively low energy. At variance with the $(p,\gamma)$ channel, for which negligible
variations with respect to previous investigations have been found, the
$(p,\alpha)$ rate has been substantially revised. The latter reaction is the main destruction channel 
of \o17~. Thanks to the significant background reduction of
the underground laboratory, a new direct determination of the 64.5 keV
resonance strength has led to the most accurate value to date, 
$\omega\gamma=10.0\pm 1.4_{stat}(\pm 0.7_{syst})$ neV. The 64.5 keV resonance dominates the reaction rate
for temperatures below 100 MK. The (bare) proton partial width of the corresponding state at
$E_X = 5672$ keV in $^{18}$F is $\Gamma_p=35\pm 5_{stat}(\pm 3_{syst})$ neV, which is about a factor of two higher
than previously claimed. As a consequence, the \rate~ reaction rate in the temperature range most relevant
for the hydrostatic H burning,
$20<T/MK<100,$ is about a factor of two higher 
than previously estimated. Larger temperatures
may be attained in explosive H burning, such as in the case of Nova events \citep{dileva2014}, and in the 
hot bottom burning (HBB) process occurring in super-AGB stars \citep{siess2010,ventura2012,doherty2014}.
For T$> 100$ MK, however, the influence of the 64.5 KeV resonance on the total
reaction rate is negligible.
In Figure 1, the corresponding variation of the  equilibrium value
of  \orat~ (Equation 1) as function of the temperature is compared
to those obtained by adopting the $^{17}$O$(p,\alpha)^{14}$N rates reported in the popular
compilations NACRE \citep{angulo1999} and STARLIB \citep{iliadis2010,
buckner2015}. As is immediately seen,
the revision of the \rate~ rate implies equilibrium values of the \orat~ ratio, 
which are up to a factor of 2 larger than those obtained so far.  

In this paper
we discuss the expected variations of the  \orat~ at the surface of stars
with different masses. In particular, we consider  the effects induced by the major deep 
mixing episodes, 
such as the dredge-up events and the extra-mixing process that is likely 
operating in bright low-mass RGB stars and, perhaps, also during the AGB phase. 
We do not address here the impact on the HBB, which is an
important nucleosynthesis episode occurring in massive AGB stars. Because of the complexity of this phenomenon, which is largely not yet fully understood, we decided to dedicate a companion paper to the discussion of the
impact  of the revised \rate~ rate on the HBB \citep{Lugaro2016}.  

A variation of the stellar \o17~ yields is also expected. According to the extant
chemical evolution models  (see, for example, the recent review of \citet{nittler2012} or 
CGE models discussed in \citet{prantzos1996,romano2003,kobayashi2011}),
 the average \o17~ abundance in the interstellar medium (ISM) should increase with time, 
mainly because of the
ejecta of intermediate and massive stars. We show that with the revision of the
\rate~ reaction rate, the \o17~ yields are substantially reduced. 
As a consequence, the rate of growth of the  \o17~ in the ISM should be
slower than previously estimated.

We computed all the stellar models presented in this paper by means
of the latest release of the full-network stellar evolution code \citep[FUNS,][]{straniero2006, piersanti2013}. 
Except for the \rate~,
all the rates of the other reactions involved in the H burning are from \citet{adelberger2011}.
When not specified, the initial composition of the models is solar \citep{lodders2009}.
It implies Z=0.014 and Y=0.27.
The various uncertainties affecting stellar models
and the related nucleosynthesis results, including those due to the 
treatment of mixing, are discussed in some detail in Section 2 of \citet{straniero2014} (see also
Sections 4 and 5 in \citet{lebzelter2015}).   
More details on the theoretical estimation of the O isotopic ratios can be found in several papers,
including \citet{dearborn1992, boothroyd1994, eleid1994, stoesz2003, karakas2014, halabi2015}.
We used the same  stellar evolution code (FUNS) as in 
\citet{cristallo2015}.  Nevertheless, 
the \o17~ abundances and yields presented in this paper
slightly differ from the corresponding values reported in the full-network repository of updated 
isotopic tables \& yields (FRUITY)
because we updated the initial composition (\citet{lodders2009} instead of \citet{lodders2003}) 
and several reaction rates, in particular, those of the CNO cycle.

\begin{figure*}
\centering
\includegraphics[width=12cm]{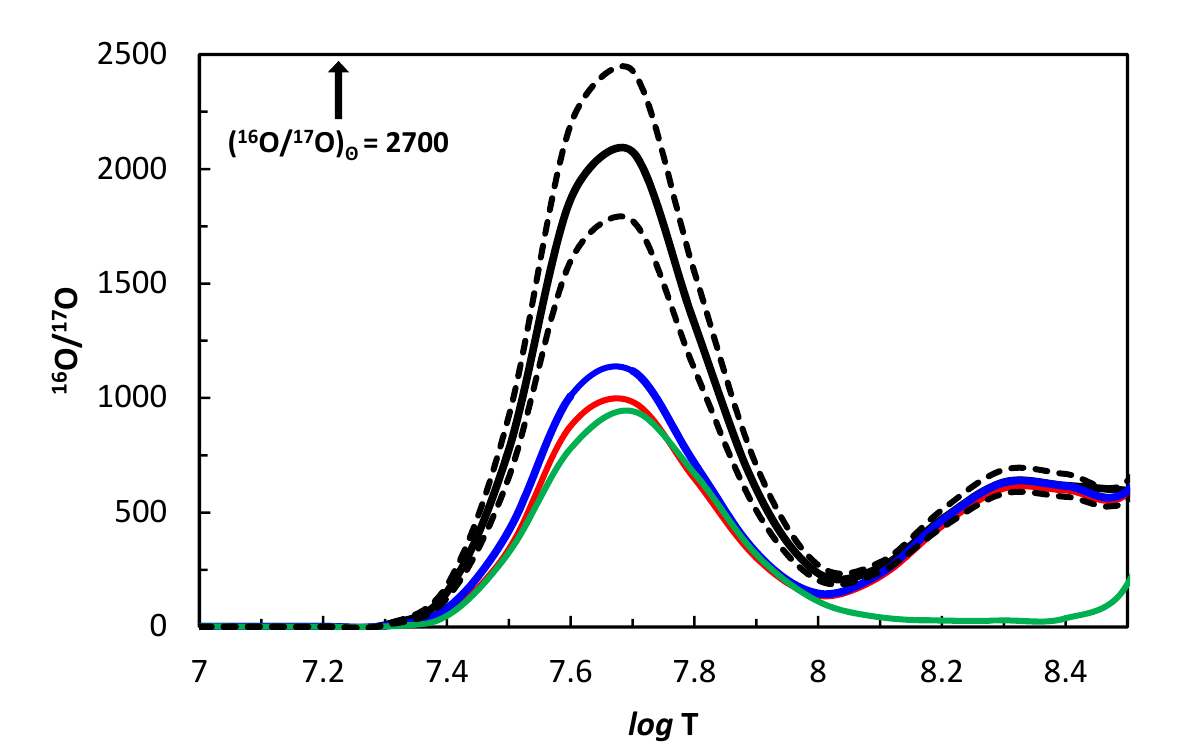}
\caption{Equilibrium values of the \orat~ vs. temperature (see equation 1). The black curve 
was obtained by adopting the revised rate of the \rate~ reaction (median values of \citet{bruno2016})
and the two black dashed lines show the recommended maximum and minimum values compatible 
with the experimental data, which corresponds to the $1\sigma$ error bar of the measured nuclear cross section.
For comparisons, the blue, red, and green lines show 
the results obtained by adopting the \rate~ rate from \citet{buckner2015}, \citet{iliadis2010}, and \citet{angulo1999}, respectively.}
\label{figura_1}
\end{figure*} 

\section{Low- and intermediate-mass stars} \label{low_mass}

In this section we discuss the variations of the \orat~ predicted by models of low- and 
intermediate-mass stars after the major dredge-up episodes, as  due to the revision of 
the \rate~ reaction rate. The possible effects on
low-mass stars undergoing RGB or AGB extra-mixing episodes are also discussed.   

\subsection{The first dredge up} \label{I_dup} 

The first appearance at the surface of a star of the chemical variations 
caused by the internal nucleosynthesis occurs 
just after the 
exhaustion of the central hydrogen when the star leaves the main sequence to become a red giant. 
A convective instability develops in the most
external layers and quickly extends down to the region where the H burning took place.  
This is the so-called first dredge up (FDU). 

The solar value of  the \orat~ is 2700
 \citep{lodders2009}.  Models of intermediate-mass stars 
($2 < M/M_\odot < 8$)
show that after the FDU this isotopic ratio is reduced down to values between
200 and 500, depending on the initial mass and composition,
 essentially because of \o17~ production.
In more detail, these stars 
develop a rather extended convective core at the onset of the core H burning, corresponding 
to the main sequence of the HR diagram.  
The central (maximum) temperature is typically between 20 and 30
 MK and the corresponding equilibrium value of the \orat~ ratio ($<100$) 
 is established in the whole convective core. 
Later on, at the epoch of the FDU,
the external convective instability penetrates  
the portion of the core that was efficiently mixed when the star was on the main sequence.
As a consequence, the surface abundance of \o17~ increases, while that of $^{16}$O slightly decreases.

\begin{figure*}
\centering
\includegraphics[width=13cm]{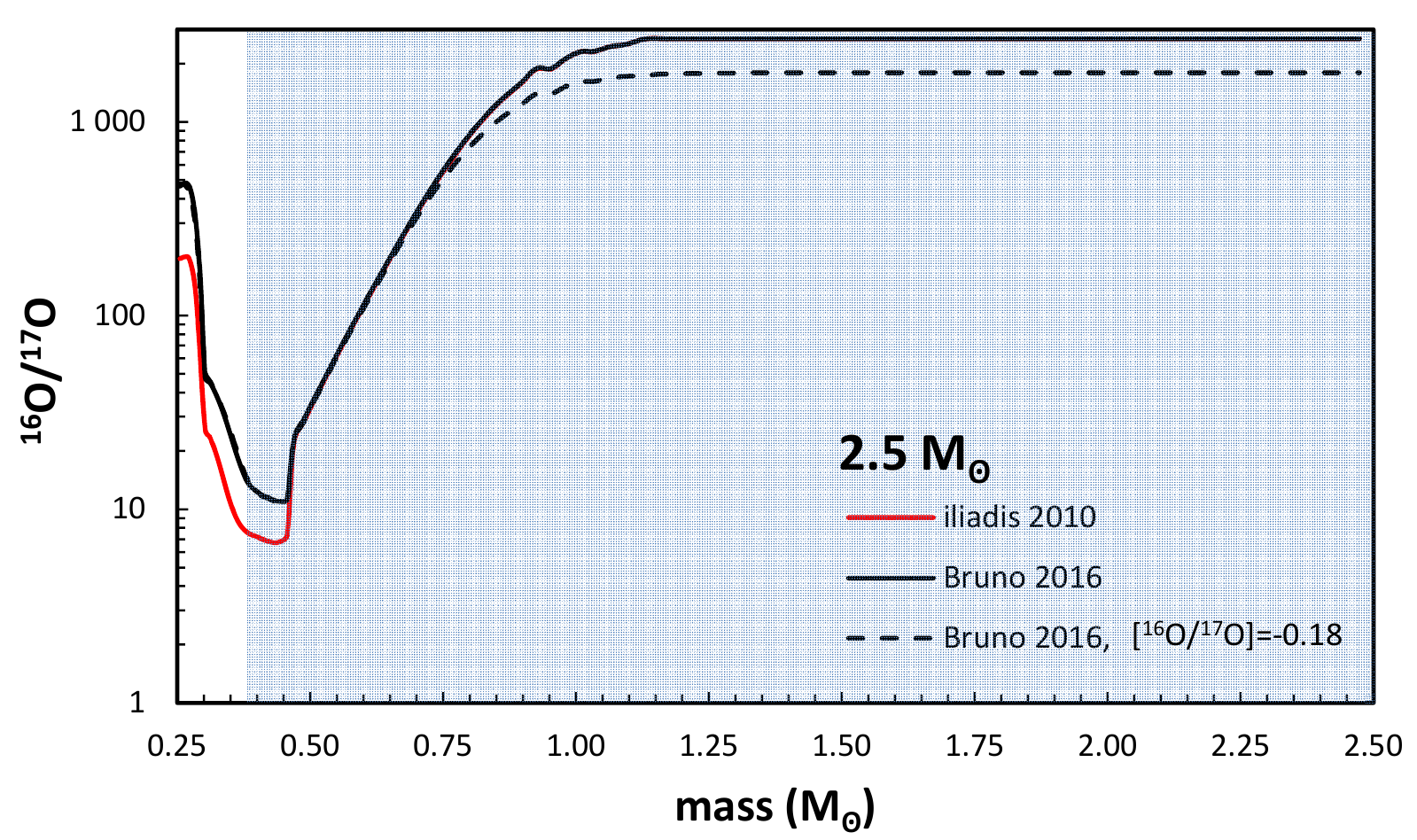}
\caption{Oxygen isotopic ratio in the H-rich envelope of a 2.5 M$_\odot$ newborn red giant star, just before the 
occurrence of the FDU. The models were computed with
different \rate~ reaction rates and/or different initial compositions (see the legend). 
The shaded area
shows the portion of the stellar envelope that are mixed at the time of the FDU.}
\label{figura2}
\end{figure*}

In Figure \ref{figura2} we plot the \orat~ profile within the H-rich envelope of a newborn giant star 
with 2.5 M$_\odot$
and solar metallicity. The three compositional profiles, which are all sampled just before the onset of the FDU, 
were obtained under
different assumptions for the \rate~ reaction rate and/or different initial composition: the \citet{iliadis2010} rate and
solar \orat~ (red line),  the new rate (median values of \citet{bruno2016}) and solar \orat~  (black line), and the new rate and
\orat~ lower than solar (black dashed line).
This figure clearly shows that in the most internal region of the envelope, where the H burning 
attained the CNO equilibrium, 
the  \orat~ does
not depend on the initial composition, but is fixed by the production and 
destruction rates of the \o17~ 
(Equation 1 and Figure \ref{figura_1}). 
 Therefore, the abundance ratio in the deepest layer of the H-rich envelope of
a giant star only depends on the reaction rates and on the maximum temperature previously 
experienced.
On the contrary, the O isotopic ratio in the external envelope
reflects the initial composition. Later on, the external convection penetrates the internal region, modifying the
surface composition. The shaded area in Figure \ref{figura2}  shows the region that is mixed by the convective 
instability at the epoch of the FDU.

\begin{figure*}
\centering
\includegraphics[width=13cm]{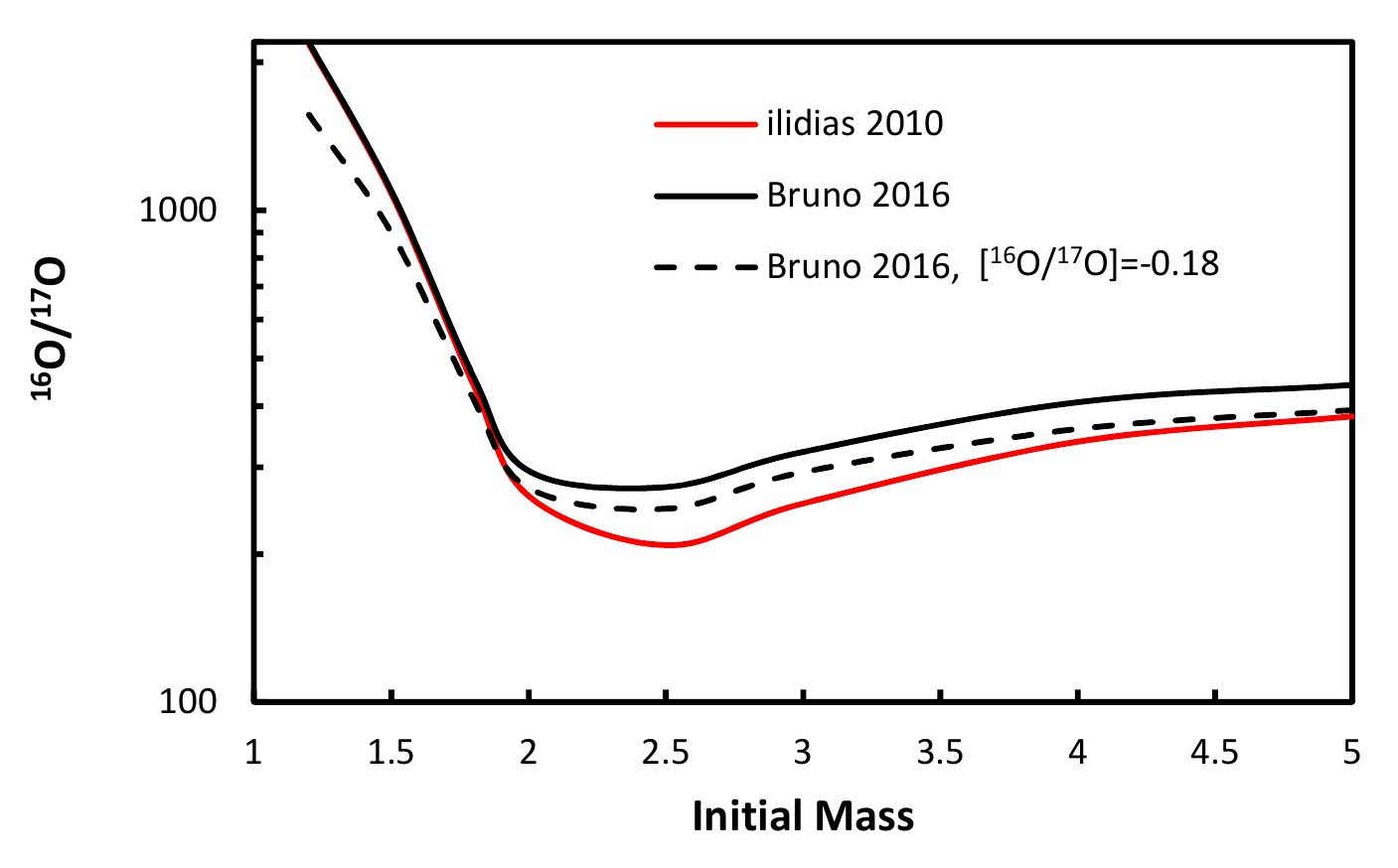}
\caption{Post-FDU \orat~ at the surface of models with different initial mass and different 
rates and/or initial composition (see the legend).}
\label{figura_3}
\end{figure*}

The maximum temperature attained in the deepest layer reached by the FDU
varies with the initial stellar mass and
composition. For a fixed chemical composition (same initial helium and metallicity), the larger the stellar mass the larger
the temperature at the innermost convective boundary. In contrast, the ashes of
the H burning are mixed with the material in the cooler external envelope with the initial composition and  
the larger the envelope mass the smaller the variation of the composition at the stellar surface.  The combination of these
two effects leads to the plots reported in Figure \ref{figura_3}, in which we show the surface value of 
the \orat~ ratio after the FDU versus the initial stellar mass. 
For convenience, the values of the post-FDU \orat~ 
are listed in Table 1.
In all cases, a sharp decrease of the \orat~
ratio is found for $1.2 < M/M_\odot < 2.5$, followed by a shallower increase for more massive stars.
In low-mass stars ($M< 1.2$ M$_\odot$), the convective core disappears at the onset of
the main sequence.
In that case the \orat~ does not change after the FDU. 

The maximum difference of 20\% between models, which are
obtained with the old and new \o17~ destruction rates, is found  
at $M=2.5$ M$_\odot$. The change implied by the new rate may be compensated by varying the initial O composition 
(dashed line in Figure \ref{figura_3}).  

\begin{table}
\label{fdu} 
\centering 
\caption{Post-FDU $^{16}$O/$^{17}$O values} 
\begin{tabular}{c c c c} 
\hline\hline 
 M/M$_\odot$ & I2010$^a$  & LUNA$^b$ & LUNA$^c$ \\
\hline
initial & 2696  & 2696 & 1797  \\
\hline
 1.2 & 2177  & 2195 & 1564  \\
 1.5 & 1081  & 1096 & 901   \\
 1.8 &  440  & 458  & 412   \\
 2.0 &  262  & 295  & 272   \\
 2.5 &  208  & 274  & 247   \\
 3.0 &  253  & 323  & 294   \\
 4.0 &  338  & 407  & 359   \\
 5.0 &  381  & 441  & 392   \\
\hline
\end{tabular}
\tablefoot{
\tablefoottext{a}{\rate~ from \cite{iliadis2010}, solar initial composition}
\tablefoottext{b}{\rate~ from \cite{bruno2016}, solar initial composition}
\tablefoottext{c}{\rate~ from \cite{bruno2016}, and initial [\orat]$=-0.18.$}
}
\end{table}

\subsection{The second and third dredge up} \label{II_IIIdup}

\begin{figure*}
\centering
\includegraphics[width=12cm]{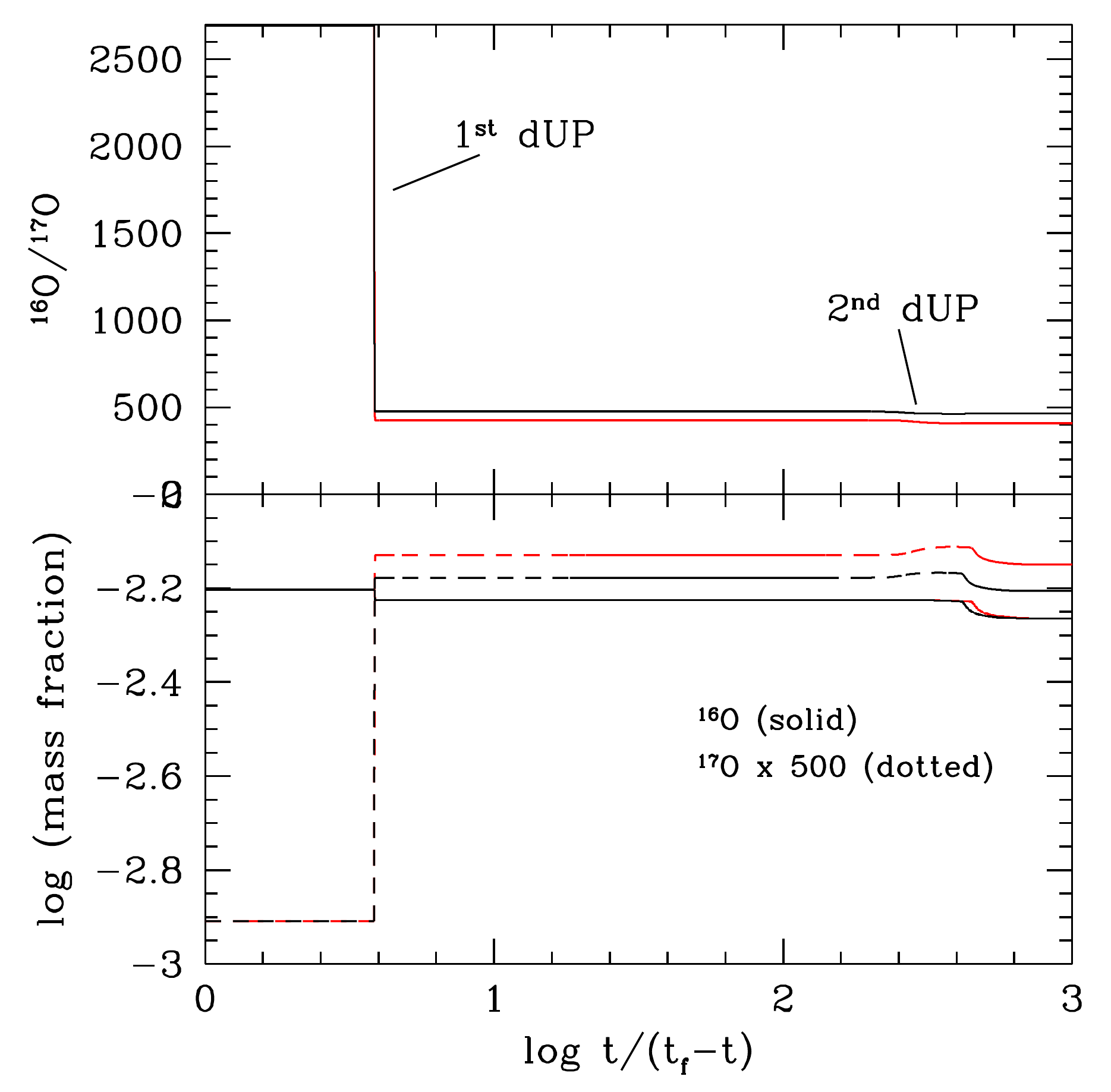}
\caption{Evolution of the two lighter oxygen isotopes in a 6 M$_\odot$ model (solar initial composition).
Upper panel: \orat~ ratio. Lower panel: $^{16}$O (solid) and \o17~ (dotted). Red lines refer to the models  
calculated by adopting the \citet{iliadis2010} rate for the \rate~ reaction, 
while black lines  represent models obtained with the new LUNA rate (Bruno et al. 2016). $t_f=68$ Myr corresponds
 to the age of the last computed model (end of the early AGB).}
\label{IIdup}
\end{figure*}

During the early-AGB phase, intermediate-mass stars ($M\geq 4$ M$_\odot$) experience a second dredge-up episode. 
This episode occurs when the
H-burning shell is  temporarily switched off and the convective envelope can penetrate  the H-exhausted core.
 In this case, H-depleted material  enriched with the ashes of the H burning is brought to the surface. 
At variance with
the FDU, the change of the \orat~ at the surface of the star is small. 
The situation is illustrated
in Figure \ref{IIdup}: in the upper panel we show the evolution  of the \orat~ at the surface 
of a 6 M$_\odot$ model (solar initial composition) up to the early-AGB phase, 
while in the lower panel we report the evolutions of the 
corresponding mass fractions of $^{16}$O (solid) and \o17~ (dotted). 
The red lines refer to the models  calculated by adopting the \citet{iliadis2010} rate for the \rate~ reaction, 
while the black lines represent models obtained with the new LUNA rate. In both cases the  variations at the
stellar surface are largely dominated by the large increase of the \o17~ abundance caused by the FDU.
Later on, the H burning restarts whereas the He-burning shell  periodically undergoes recursive thermonuclear runaways 
(i.e., thermal pulses; TP). Then, third-dredge-up episodes occur 
after each thermal pulse, 
except for the first and a few final TPs. 
As for the second dredge up, the convective envelope penetrates the H-exhausted core. 
However, the variation of the O 
isotopic ratio caused by the third dredge up is generally small. 
An increase of the \orat~ is actually found in low metallicity models owing to the dredge up of primary 
$^{16}$O produced by the He burning \citep[see, for example, 
O abundances listed in the FRUITY database][]{cristallo2015}, but this phenomenon is 
independent of the adopted CNO reaction rates.

\subsection{Extra mixing in RGB and AGB stars} \label{emix}

\begin{figure*}
\centering
\includegraphics[width=12cm]{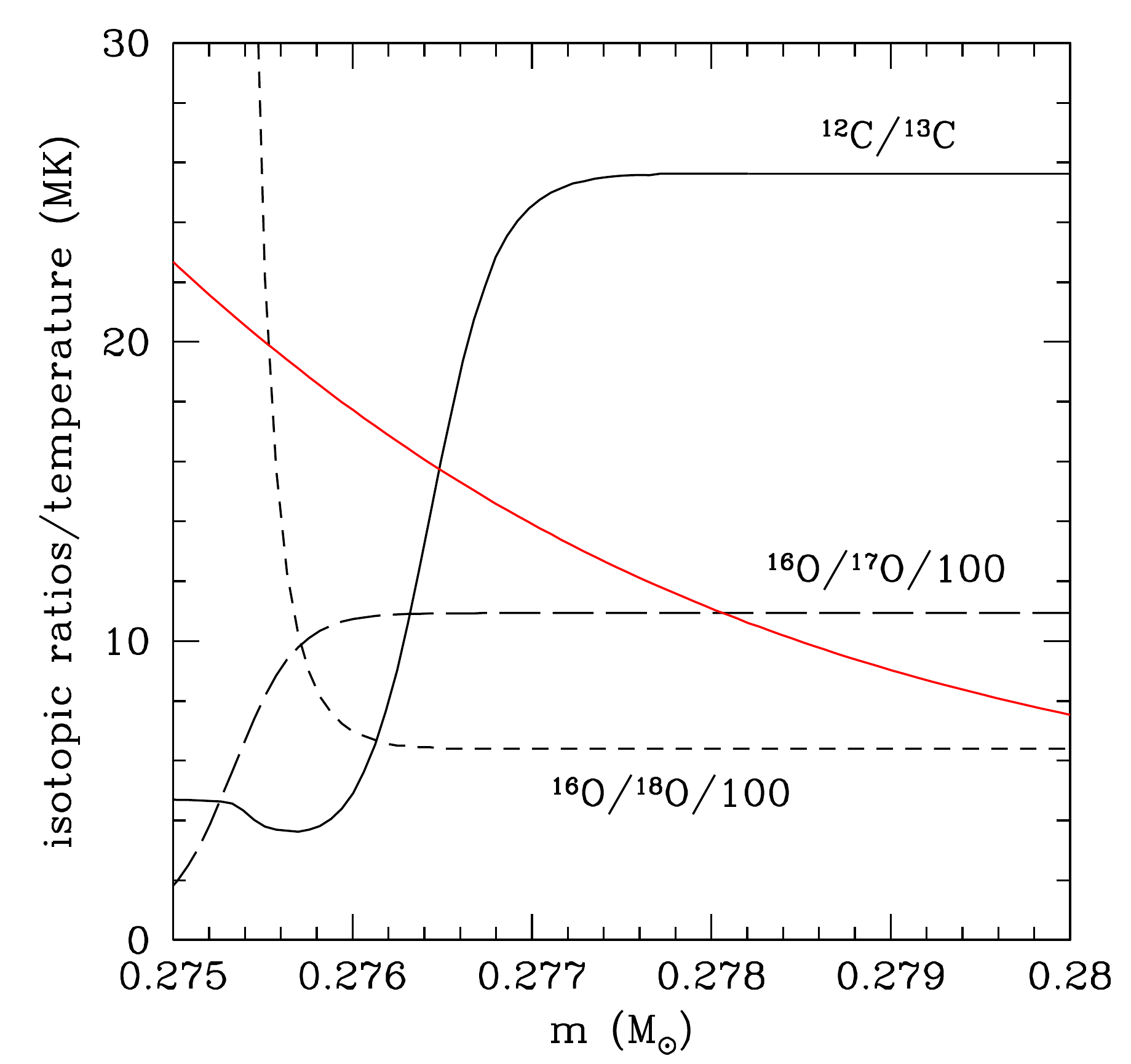}
\caption{This figure shows the C and O isotopic ratios in the innermost portion of the
H-rich envelope of a RGB-bump model with $M=1.5$ M$_\odot$ (solar initial composition):
$^{12}$C/$^{13}$C (black solid line), $^{16}$O/$^{17}$O (short dashed line), and $^{16}$O/$^{18}$O (long dashed line).
The red line represents the temperature profile (in MK).}
\label{1p5prof}
\end{figure*}

It is generally accepted that the FDU alone cannot account for  the CNO abundances measured in the brightest RGB stars with
$M<2$ M$_\odot$  and that an extra-mixing process should be at work in the upper RGB 
\citep{charbonnel1995}. In particular, the $^{12}$C/$^{13}$C ratio in these stars is substantially lower than predicted 
by stellar models after the occurrence of the FDU.
On the contrary, model predictions are in good agreement with the C isotopic ratios measured in more massive RGB stars 
as well as in faint RGB stars with low mass. 
For instance, values of $^{12}$C/$^{13}$C$<20$ are commonly found in low-mass RGB stars 
that are brighter than the so-called RGB bump, i.e., the peak in the RGB 
luminosity function due to the temporary stop of the luminosity growth that occurs when 
the H-burning shell approaches the 
chemical discontinuity left by the FDU. 
  \citep{gilroy1991,gratton2000,abia2012}. In
contrast, model predictions for  the  $^{12}$C/$^{13}$C ratio after the FDU are always $> 20$. 
In principle, the O
isotopic ratios may also be affected by the extra mixing, although there is some observational evidence 
against such a possibility \citep{abia2012}.  
In this context, the new (stronger) rate of the \rate~ reaction can affect the variation of the 
\o17~ at the stellar surface. In Figure \ref{1p5prof} we report C and O isotopic ratios 
in the innermost layers of the
H-rich envelope of a RGB-bump model with initial mass 1.5 M$_\odot$ and initial solar composition. 
Sizeble variations of C, N, and O isotopes, as due to H burning, occur at $m\le0.277$ M$_\odot$.
Since the internal boundary of the convective envelope is at
$m=0.292$ M$_\odot$, extra mixing is needed to bring the product of the CNO nucleosynthesis 
into the convective zone. In the same figure, the temperature profile is also reported (red line).  
A modification of the \orat~ ratio would require a much deeper extra mixing than that needed 
to explain the C isotopic ratio commonly observed in bright low-mass RGB stars. 

\begin{figure*}
\centering
\includegraphics[width=12cm]{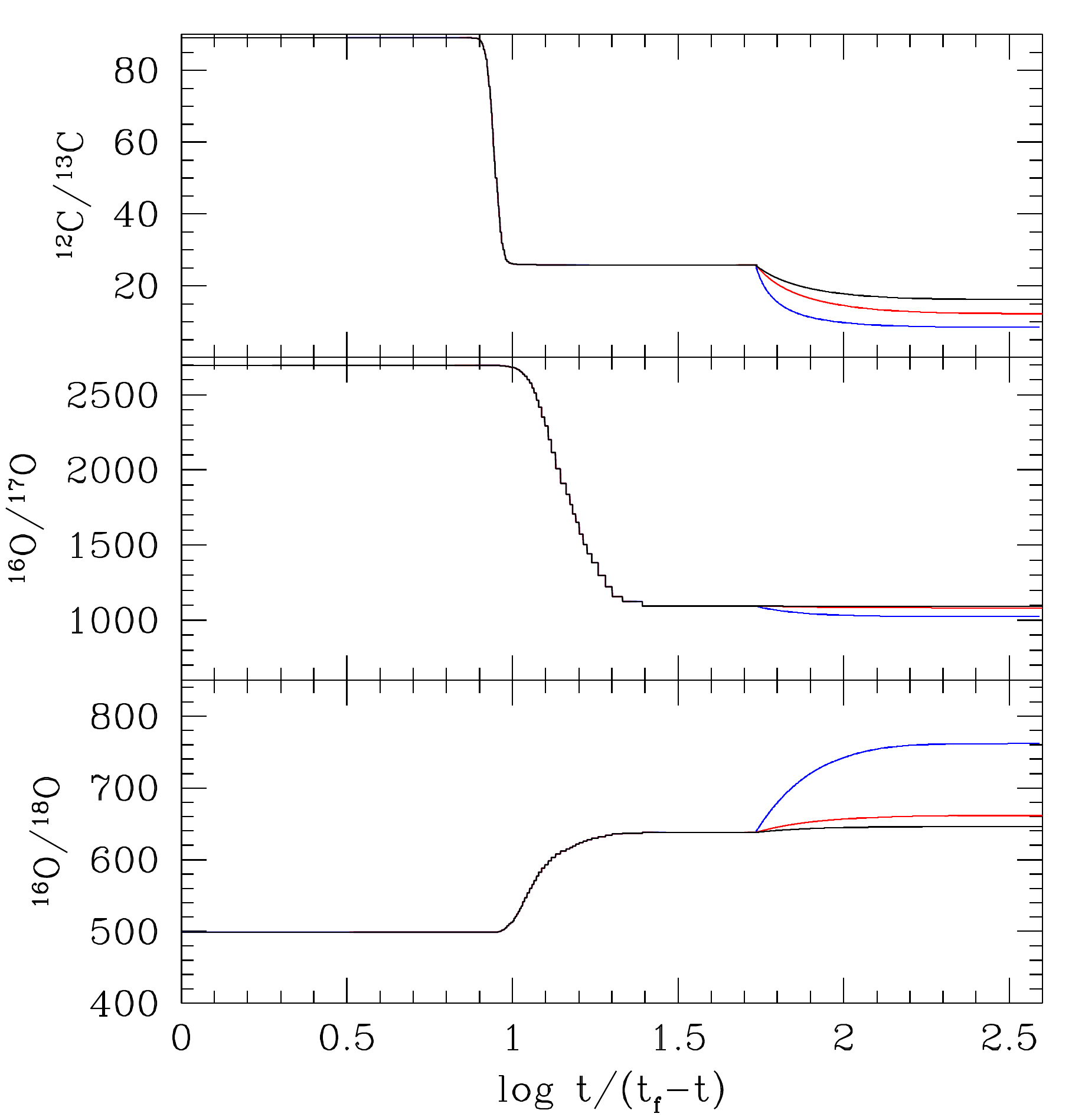}
\caption{Evolution up to the RGB tip of C and O isotopic ratios in the 3 models with extra mixing described in the text.
The maximum temperature attained by extra mixing is T$_{max}$ = 20, 21, and 23 MK (black, 
red, and blue lines, respectively). In all cases, the fluid velocity
in the extra-mixing zone is about 1/100 of that in the fully convective envelope. 
The evolutionary timescale has been properly rescaled to
$t_f = 2.91$ Gyr, which corresponds to the onset of the core-He burning. 
The FDU starts at $\log t/(t-t_f) \sim 1$, while  the extra-mixing process is activated 
just after $\log t/(t-t_f) \sim 1.7$, namely, the epoch of the RGB bump.}
\label{extrares}
\end{figure*}

Since the actual physical process that is capable of driving an efficient mixing 
below the convective envelope in giant stars is still largely unknown, 
a few free parameters are usually introduced to investigate the effects of these phenomena.  
We do not pretend to discuss these effects or even the nature of the extra-mixing processes.
Several paper have been already dedicated to this subject 
\citep[see,e.g.,][and references therein]{dearborn1992, boothroyd1994,
eleid1994, charbonnel1995, nollet2003, denissenkov2003, busso2007, eggleton2008, charbonnel2010,
palmerini2011}.
The models presented here just aim to highlight possible
effects of the revised \rate~ rate. 
Starting from the RGB-bump model shown in Figure \ref{1p5prof}, 
we switched on an artificial extra-mixing
process extending from the convective boundary down to a maximum temperature T$_{max}$. 
We set the mixing velocity to 100
cm/s, which is about 2 orders of magnitude slower than that in the convective region. 
Mixing velocities of this order of magnitude are expected in the case of 
thermohaline mixing \citep{charbonnel2007} and magnetic buoyancy \citep{palmerini2009}, while 
somewhat lower values have been estimated for rotation induced mixing \citep{palacios2006}.
We verified that the C and O isotopic ratios at the stellar surface 
does not substantially depend on the assumed mixing velocity, but depends mainly on T$_{max}$. 
In Figure \ref{extrares} 
we show the evolution of the C and O
isotopic ratios for T$_{max}$ = 20, 21, and 23 MK. The most extreme case implies an
extra-mixing zone extending from the internal convective boundary down to the outer border of the H-exhausted core.
After the activation of the extra mixing, in all three cases the
$^{12}$C/$^{13}$C quickly drops below 20. The resulting C isotopic ratios 
are in good agreement with the observed values for the brightest low-mass RGB stars. 
In contrast, one
only finds a non-negligible enhancement of the $^{16}$O/$^{18}$O in the most extreme case, whilst in spite of the stronger rate of the
\rate~ reaction,  the $^{16}$O/$^{17}$O practically remains unchanged.

\begin{figure*}
\centering
\includegraphics[width=12cm]{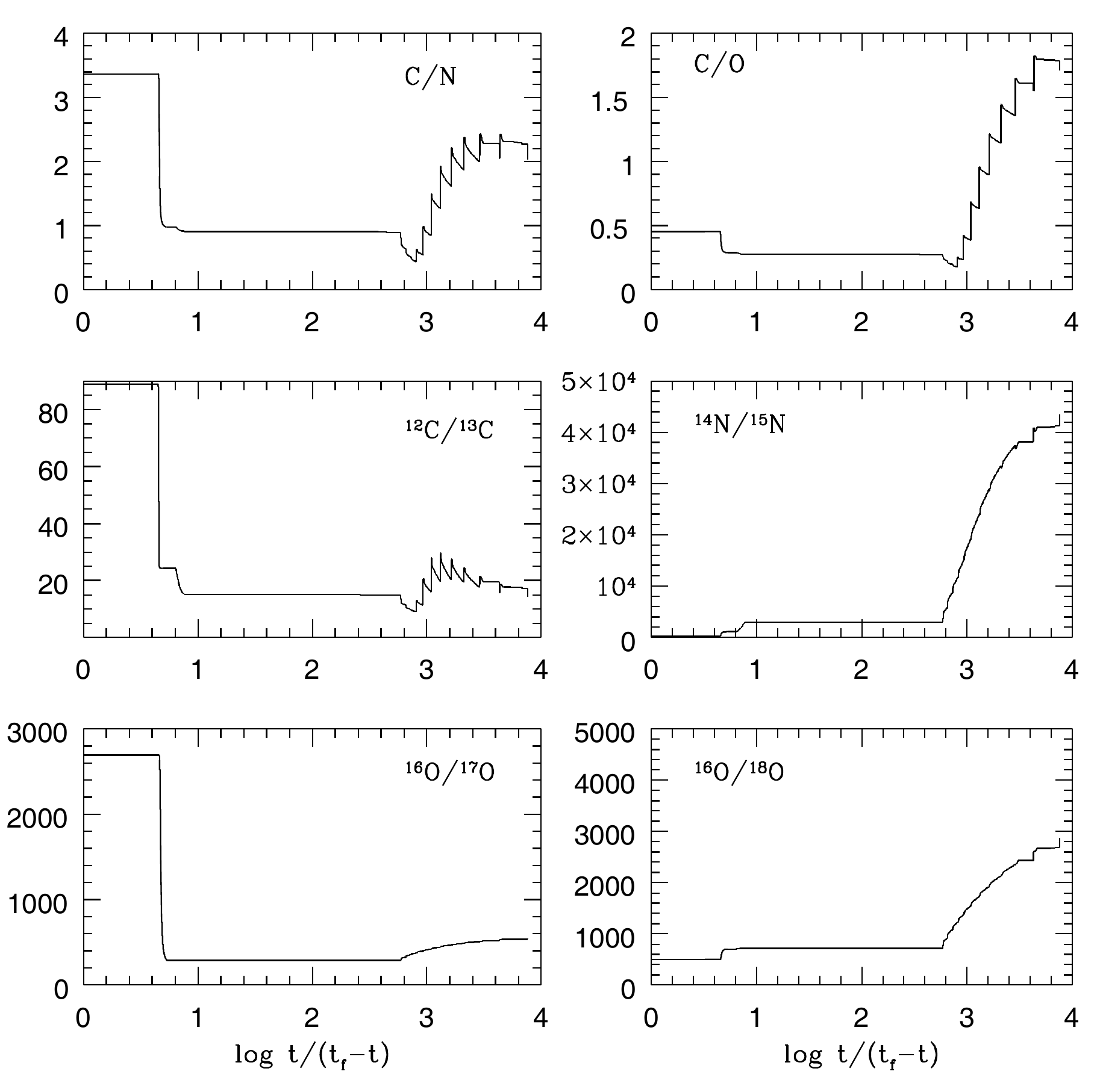}
\caption{Evolution up to the AGB tip of C, N, and O elemental and isotopic ratios in a 2 M$_\odot$ model
with both RGB and AGB extra mixing. In the first case the maximum temperature was fixed to T$_{max} = 22$ MK, 
while in the latter T$_{max} = 40$ MK (see text for further details). The evolutionary timescale was rescaled to 
$t_f = 1.196$ Gyr, which corresponds to the age of the last computed model (AGB tip). 
The FDU occurs at $\log t/(t-t_f) \sim 0.7$, while some effect of the RGB extra mixing is visible a bit later. 
Finally the combined actions of the  and the AGB extra mixing arise at  $\log t/(t-t_f) > 2.8$.}
\label{agbextra}
\end{figure*}

In principle, even hotter regions may be attained by an AGB extra mixing. However,  the actual 
occurrence of this process is still largely debated \citep[see, for example,][]{lebzelter2008,
lederer2009,karakas2010,busso2010,stancliffe2010}. Nevertheless, in order to investigate the effects of 
the revised rate of the \rate~ reaction on the O isotopes of AGB stars possibly undergoing 
extra mixing, we extended the computation of the 2 M$_\odot$ model up to the AGB tip.  
An AGB extra mixing extending down to T$_{max}=40$ MK was activated since the first termal pulse. 
In addition, a RGB extra mixing (T$_{max}=22$ MK)
was also activated after the occurrence of the RGB bump. In both cases the average fluid velocity in the
 extra-mixing zone was set to 100 cm/s. 
The result is
shown in Figure \ref{agbextra}. The $^{16}$O/$^{17}$O should increase in the case of an AGB extra mixing, contrary to the variations induced by the
FDU and, to a lesser extent, by the RGB extra mixing. The reason for this increase is simple: after the FDU, the
$^{16}$O/$^{17}$O drops down to $\sim 300$, while at T=40 MK the equilibrium value of this 
ratio is $\sim 1870$ (see Figure 1). Therefore, as a consequence of the AGB extra mixing, the material with high \orat~ 
of the most internal portion of the H-burning shell is mixed with the envelope material that is characterized by 
low \orat~. This effect is weaker
for stars with smaller initial mass because the \orat~ after the FDU is higher (see Figure 3).
With previous \rate~ reaction rates (e.g., Iliadis et al. 2010), the equilibrium 
values of the \orat~ would be a factor of two smaller, thus reducing the overall increase of this isotopic 
ratio at the stellar surface. 
As a whole, in spite of the 
larger \rate~ rate, the variation of $^{16}$O/$^{17}$O 
caused by the AGB extra mixing is much smaller than those of the other CNO isotopic ratios, such as $^{12}$C/$^{13}$C,
$^{14}$N/$^{15}$N, and $^{16}$O/$^{18}$O. At the beginning of the C-star phase, i.e., when C/O=1, 
$^{16}$O/$^{17}$O is $ \sim 400$, which is a 30\% larger than the value obtained with 
the rate provided by \citet{iliadis2010}. We note that C stars affected by AGB extra mixing should 
show extreme values of $^{14}$N/$^{15}$N ($> 2\times 10^4$) and $^{16}$O/$^{18}$O ($>2\times 10^3$),
while the  $^{12}$C/$^{13}$C should be $\le 20$.  

\section{Yields} \label{massive}

Oxygen-17 is a secondary product of the H burning and, therefore,  
all stars contribute to its synthesis.  
The major contribution to the pollution of the interstellar medium likely comes from massive stars ($M > 10$ M$_\odot$).
The winds of red giant stars of intermediate and low mass provide further contributions on a longer timescale. 
Extant chemical evolution models show that the
abundance of \o17~ in the ISM of the galactic disk increases with time 
\citep[see the review of][and references therein]{nittler2012}. 
This expectation is substantially confirmed by the abundance analysis  of stars belonging 
to stellar populations of different ages and in the ISM \citep[see, e.g.,][]{nittler2012, lebzelter2015}.
The \o17~ increase with time is stronger toward the Galactic center than in the outermost regions because of the increase of the stellar density at short Galactic radius and the
consequent increase of the pollution rate \citep[see figures 1 and 2 in ][]{nittler2012}.  
Extant nucleosynthesis models show that stars with mass between 2 and  20 M$_\odot$ should 
release a few $10^{-5}$ M$_\odot$ of \o17~.  
Because  \o17~  is a secondary isotope,
its production is higher at larger metallicity. More precisely, the yield depends on the initial C+N+O 
abundance.  

As for intermediate-mass stars, in models of massive stars
 the amount of \o17~ in the envelope is also enhanced as a consequence of the FDU. 
Here, we refer to the progenitors of type II supernova, 
which retain a H-rich envelope up to the final explosion.
 On the contrary, the \o17~ is destroyed
in the H-exhausted core. Then, the \o17~ in the envelope remains unchanged after the FDU and 
until the core collapse. Also the final explosion does not affect  
the \o17~ yield; see, e.g., \citet{limongi2000,limongi2001}, where both 
pre- and post-explosive yields are discussed, or \citet{kobayashi2011}, where it is shown that 
the  \o17~ yield does not depend on the explosion energy. Indeed,
the H-rich envelope expands prior to the passage of the shock wave,
so that the explosive nucleosynthesis is inhibited. 
In practice, a reasonable estimation of the \o17~  yield 
could be obtained by multiplying the \o17~  mass fraction after the FDU by the total mass of 
H-rich material ejected during the long hydrostatic evolution (by stellar wind) 
or as a consequence of the final explosion. Therefore, 
the adopted mass loss rate after the FDU, and the details of the final explosion, are not 
critical to modeling the \o17~ yield in massive stars. This approximation is reasonable for most 
massive stars, except in the case of an early ejection of the whole envelope that stops the growth 
of the H-exhausted core, as occurs in Wolf-Rayet stars.      
In contrast, there is a major uncertainty 
because of the formation of a semi-convective shell instability during the core H burning, when the
fully convective core recedes and leaves a decreasing H profile.  As is well known, 
depending on the adopted mixing scheme, 
the core He burning may start when the star is still a blue compact structure or after it has expanded 
and become a red giant \citep[see, e.g.,][figure 3, in particular]{langer1995}. In the former case, the occurrence of the FDU is delayed; more precisely, 
it occurs after the end of the core-He burning. In this case the amount of H-rich material lost
before the FDU may be greater and, therefore, the total yield is smaller.

In the models of massive stars presented here, we adopted a semi-convective scheme, i.e.,
the degree of mixing is calibrated in order to establish convective neutrality\footnote{
$\nabla_{adiabatic}-\nabla_{radiative}=0$, where $\nabla=\frac{\partial \ln T}{\partial \ln P}$.}
in the whole 
semi-convective layer. In this case the mixing efficiency is substantially limited, so that 
the FDU occurs soon after the main-sequence phase. As a result, most of the envelope mass is lost
after the occurrence of the FDU. Then, we computed three models 
with solar initial composition and masses
12, 15, and 20 M$_\odot$, which are representative of the most common type II supernova 
progenitors.
We stopped the computation at the end of the C-burning phase. At that time, 
the residual stellar lifetime is so short that the composition of the 
H-rich envelope and the H-exhausted core mass remain frozen up to the final explosion. 
Therefore, we obtained the total yield by adding the \o17~ mass contained in the 
H-rich envelope of the last computed model to the amount of \o17~ cumulatively ejected before
the end of the C-burning phase.       

The resulting  yields are reported in Table 2; they  are net yields, namely
the difference between the  \o17~ mass ejected during the whole stellar lifetime and the
initial total \o17~ mass. Results for different \rate~ reaction rates are provided in columns 2 and 3.
Once the different initial compositions and reaction rate libraries 
are considered, 
the yields listed in column 2  are in good agreement
with those reported by \citet{kobayashi2011} and \citet{chieffi2013} (non-rotating models). 
Yields of a selected sample of low- 
and intermediate-mass stellar models with solar composition, as described in the previous sections, are 
reported as well. In this case all the models were evolved up to the AGB tip and no extra-mixing
processes were included.  

The new LUNA rate implies a sizeable reduction of the net yield of \o17~of up to a 40\%. 
This result is almost independent of the initial metallicity or He content. 
The impact on models of Galactic chemical evolution deserves further investigation. 
In particular, we expect a
slower rate of \o17~ pollution of the ISM (see the discussion in section 4).      

\begin{table}
\label{yields} 
\centering 
\caption{$^{17}$O yields$^a$} 
\begin{tabular}{c c c c} 
\hline\hline 
 M & I2010$^b$  & LUNA$^c$ &  var(\%) \\
\hline
1.5 & 0.16  & 0.17  &  -1 \\
2.5 & 4.78  & 3.57  &  -22 \\
  4 & 5.08  & 4.19  &  -15 \\    
  6 & 5.70  & 4.70  &  -17 \\
 12 & 5.95  & 4.70  &  -21 \\
 15 & 4.42  & 3.52  &  -20 \\
 20 & 2.38  & 1.38  &  -42 \\
\hline
\end{tabular}
\tablefoot{
\tablefoottext{a}{Net yields are listed in $10^{-5}$ M$_\odot$}
\tablefoottext{b}{\rate~ from \cite{iliadis2010}, solar initial composition}
\tablefoottext{c}{\rate~ from \cite{bruno2016}, and solar initial composition.}
}
\end{table}
 
\section{Conclusions}

We have revised the theoretical predictions of the \orat~ ratios and of the \o17~ 
yields from stars in a wide range of initial masses.
We based all the computations on the new recommended values of
the  \rate~ reaction rate as described by \citet{bruno2016}. \citet{bruno2016} also provide
an upper and a lower limit for this reaction. As seen in Figure 1, the \orat~ values found using 
rates from previous 
compilations are outside the updated error band. 
Similarly, the uncertainty implied for the \o17~ yields is much smaller than the 
difference between the revised values and the previous evaluations.

Reaction rates with uncertainties at the level of a 20\% (or lower), 
as recently achieved by the LUNA collaboration for 
the \rate~ and $^{17}$O$(p,\gamma)^{18}$F reactions, are required to constrain 
the efficiency of the various deep-mixing
episodes operating in stars of various masses. 
Thanks to the improved accuracy of the relevant reaction rates, it will become possible 
to check the reliability of the numerical algorithms used to model stellar convection 
and other dynamical and secular instabilities in stellar interiors. In this 
context, comparisons of model predictions with \orat~ ratios in giant stars is of primary importance, 
providing calibration tools for the various free parameters whose values are not fixed by the theory. 
Similarly, O-rich, pre-solar grains may also provide useful constraints for the models 
\citep{Lugaro2016}.
The revision of the \o17~ yields could have a major impact on Galactic chemical evolution studies. 
In particular it should help the interpretation of the oxygen isotopic composition of various 
Galactic components, such the Sun, giant stars, and diffused matter. In this context,  
\citet{lebzelter2015}, analyzing post-FDU measurements of the O isotopic 
ratios in RGB  stars belonging to Galactic
open clusters, conclude that the O isotopic composition of the Sun does not represent 
the original composition of the gas from which these stars formed. 
In particular, they were able to reproduce the observed O abundance by assuming
smaller-than-solar initial O isotopic ratios.
Possibly, this occurrence may be explained in the framework of
Galactic chemical evolution. Indeed, as shown by, for example, \citet{prantzos1996,romano2003,
kobayashi2011,nittler2012},
the \orat~ ratio decreases with time, so that stars that are younger than the Sun should have been 
formed, on the average, from a gas characterized by smaller oxygen isotopic ratios. 
Models obtained with the new and higher rate of the \rate~ reaction, 
which implies larger \orat~ after the FDU,  
confirm and even reinforce the need of smaller-than-solar initial \orat~ to reproduce 
the ratios observed in post-FDU 
giants belonging to Galactic open clusters. For instance, we obtained the models represented with dashed 
lines in both Figs. 2 and 3 by
assuming an initial \orat=1797; this value is lower than the solar value (2700) and roughly 
corresponds to the weighted average \orat~ ratio of the interstellar medium in the 
region of the Galactic disk closer to the Sun (between 6 and 10 Kpc from the Galactic center), 
as derived from measurements of CO molecular lines  
(\citet{Wouterloot2008}; see also \citet{penzias1981}. On the other hand, \citet{kobayashi2011}
found that their GCE model underestimates the  solar \orat~ and conclude that a reduction 
of the \o17~ yields could solve this problem. As shown in section 3, the revision of the
\rate~ reaction rate goes in this direction and, therefore, it could alleviate the problem.   

As a whole, more stringent constraints to important phenomena, 
such as stellar migration or the observed composition gradients along the Galactic disk, 
and to the nucleosynthesis in general are expected by comparing the observed abundances of  
oxygen isotopes with predictions of chemical evolution models based on 
more accurate oxygen yields.

\begin{acknowledgements}
We are indebted to Maria Lugaro and Paola Marigo for their careful reading of the manuscript, 
comments, and suggestions.
We thank Thomas Lebzelter, Carlos Abia, Inma Dominguez, and Marco Limongi for the many helpful discussions about 
 \orat~ in RGB, AGB, and massive stars. 
\end{acknowledgements}

\bibliographystyle{aa} 
\bibliography{oiso_3}

\end{document}